\documentclass[a4paper,12pt,epsf,amssymb,citesort]{article}

\usepackage{tabularx}
\usepackage{array}
\usepackage{graphics}
\usepackage{graphicx}
\usepackage{epsfig}
\usepackage{amsmath}
\usepackage{amssymb}
\usepackage{cite}
\usepackage{verbatim}

\def\lsim{\raise0.3ex\hbox{$<$\kern-0.75em\raise-1.1ex\hbox{$\sim$}}}
\def\gsim{\raise0.3ex\hbox{$>$\kern-0.75em\raise-1.1ex\hbox{$\sim$}}}
\def\noi{\noindent}  \def\bea{\begin{eqnarray}}
\def\eea{\end{eqnarray}} \def\beq{\begin{equation}}
\def\eeq{\end{equation}} 
\def\beeq{\begin{eqnarray}} \def\eeeq{\end{eqnarray}} \def\R{ {\rm R
\kern -.31cm I \kern .15cm}} \def\C{ {\rm C \kern -.15cm \vrule
width.5pt \kern .12cm}} \def\Z{ {\rm Z \kern -.27cm \angle \kern
.02cm}} \def\N{ {\rm N \kern -.26cm \vrule width.4pt \kern .10cm}}
\def\1{{\rm 1\mskip-4.5mu l} }
\newcommand{\cb}{\mathcal{B}}

\begin{document}
\title{\bf Unexpected pattern of transitions to radial excitations by heavy quark current}
\author{\scalebox{.92}{\sc D.~Be\v{c}irevi\'c, A.~Le~Yaouanc, L.~Oliver and J.-C.~Raynal}}
\maketitle

\begin{center}
{\sl Laboratoire de Physique Th\'eorique (B\^at. 210)} \footnote{Unit\'e Mixte de Recherche UMR 8627 - CNRS}
\vskip 1.6 truemm
{\sl Univ. Paris-Sud, Universit\'e de Paris-Saclay, 91405 Orsay, France.}
\end{center}

\begin{abstract}
Motivated by the claimed possibility of a large contribution of the first radial excitation of the $D^{(\ast )}$ to the $B$ semileptonic decay into charmed mesons, also invoked to solve the ``$1/2$ vs. $3/2$ semileptonic puzzle", we discuss the transitions to heavy-light radial excitations by a heavy $b \to c$ quark current. We first consider a HQET sum rule, which provides a bound on the slopes of Isgur-Wise functions which we then calculate in the Bakamjian-Thomas framework which both guaranties covariance in the heavy quark limit and satisfies a set of HQET sum rules. We observe a remarkable property that for a large variety of wave functions the transition to the first radial excitation is very small while the transition matrix element to the second radial excitation is large and dominant in saturating the HQET sum rule. This is opposite to what is found in non-relativistic models, where the transition to the first radial excitation dominates the sum rule. 
The relative magnitude of the transition to the second excitation appears to be weakly dependent on the dynamical scale (radius of the bound states), and the same holds true for the slope of the elastic transition. These features could be tested in the heavy mass limit of lattice QCD. This pattern is shown to be related to the general structure of the Bakamjian-Thomas model, it is independent of the spin structure of the approach and derives mainly from the Lorentz transformation of the spatial wave function, a feature often disregarded in quark models. 

\end{abstract}

\noi {\small LPT-Orsay-17-22}

\vskip 0.5cm

\noi {\footnotesize Damir.Becirevic@th.u-psud.fr}\par
\noi {\footnotesize Alain.Le-Yaouanc@th.u-psud.fr}\par
\noi {\footnotesize Luis.Oliver@th.u-psud.fr}

\section{Introduction}
\vskip 0.5cm
\subsection{Motivation}

Large discrepancies have been observed when comparing theoretical predictions for $\cb (B \to D^{**}(L=1) \ell \nu )$ with the data~\cite{memorinos}. 

Radial excitations $D^{{(n)}}(L=0)$ could constitute a background to $B \to D^{**} (L=1) \ell \nu $, since they too can decay to $D \pi$ or $D \pi \pi$. 
A large contribution from the first one has been invoked to explain the ``$1/2$ vs. $3/2$ puzzle" \cite{bernlochner}, using both a QCD sum rule approach, and a quark model result from Ref.~\cite{galkin1}. The resulting rate is $\cb(B \to D^{(*)'} \ell \nu)  \approx {\cal O} (1\%)$. Note that only the first radial excitation can be relevant in the semileptonic decay, because the excitation energy $\Delta E$ is otherwise too large for a decay rate to be sizeable. 

From now on, we shall denote by $\frac{d} {dw } \xi^{(n)}(w)\big|_{w=1}\equiv \frac{d} {dw } \xi^{(n)}(1)$  
the derivative of the Isgur-Wise (IW) function at zero recoil, for a transition to the $n = 1 ,2,...$ radial excitation ($n=0$ being the ground state). 
The value obtained in Ref.~\cite{galkin1}, $\frac{d} {dw } \xi^{(n=1)}(1)=2.2$, indeed allows for a large $\cb(B \to D' \ell \nu )$.

\subsection{Methods to evaluate transition amplitudes} \label{methods}.

It is useful to examine whether or not various models or approaches in the heavy quark limit yield the same order of magnitude for $\frac{d} {dw } \xi^{(n=1)}(1)$.

\begin{enumerate}

\item HQET sum rules

There exists a large number of Heavy Quark Effective Theory (HQET) sum rules which severely constrain the properties of the IW functions for transitions to excited states, including the radial ones. They derive from the Operator Product Expansion (OPE), and they provide bounds that can also be established in the Lorentz-group  approach to IW functions~\cite{Lorentz}. They can be used to check the consistency of concrete calculations and models in the infinite mass limit, and we shall use this possibility for the present problem, cf. Sec. 2. The sum rule that we have established in Ref.~\cite{SRradialHQET} is especially relevant here as it provides an explicit bound to the slopes of IW functions. 

\vskip 0.2cm
\item Lattice QCD approach

This is, in principle, the only method of calculation based on first theory principles and it has been successfully used to study many hadronic transitions. However, it is not easy to use lattices in the present problem. Studying masses of radial excitations is already difficult as standard methods give reliable estimates of the lowest state of a given $J^P$, and sophisticated methods should be used to extract the higher states contributing to the correlation functions. Extracting the inelastic transition matrix elements is even more challenging.~\footnote{A method of choice to extract the properties of radially excited states on the lattice is known as {\sl Generalized Eigenvalue problem} ``GEVP" \cite{GEVP}.}  Despite these difficulties it is nowadays possible to perform trustable computations of hadronic matrix elements involving the first excited states, such as the decay constants, hadron-to-hadron transitions via the light quark current in the heavy quark mass limit, or the electromagnetic transitions~\cite{transitions-to-radial}.
For the transitions through a heavy quark current, in which we are interested in this paper, a first attempt to compute IW functions has been made a long time ago in Ref.~\cite{hein} in the limit of infinitely heavy both $b$ and $c$ quarks, by using non-relativistic QCD (NRQCD) on the lattice. This work was preliminary and studied only the transition to the first radial excitation, while in fact we also need the IW function to the second one. 
\vskip 0.2cm

\item Lattice QCD and quark models

Note that lattice QCD may also be useful indirectly to assess the validity of the quark models for the calculation of the IW functions that we are using below. 
For instance, one can check the relevance of potentials and wave functions by comparison with lattice data in the heavy quark limit of QCD. 
This is what we have done for the spatial distributions of current densities in the static limit in both approaches, or for the coupling constants corresponding to integrated 
distributions, cf. Ref.~\cite{spatial}. A comparison with quark models has also been made for the transitions to radial excitations induced by the light quark current 
in the static limit in Ref.~\cite{blossierDistr}.
\vskip 0.2cm 

\item Quark models

Although inherently approximate the quark model is a tool to formulate quick and definite predictions for a large set of hadronic excited states, 
and for a large range of the kinematical variables. However, there is a large variety of quark models, and one has to take into account both their 
theoretical soundness and their phenomenological achievements before applying them to a given physical situation. 

\begin{itemize}

\item There are several general frameworks for treating bound states, such as the Schwinger-Dyson and Bethe-Salpeter formalisms~\cite{roberts}, or the quasipotential approach of Ref.~\cite{sadzjian}, which is also at the origin of the works of Refs.~\cite{galkin1,faustov}, which have been applied to a large number of processes. Another possibility is the family of models based on the general Bakamjian-Thomas (BT) framework. For the latter, we have different forms of the dynamics: the instant form, like the one we choose below [see also Ref.~\cite{covariant}], the light-front or null-plane form~\cite{simula} (or, almost equivalently, the dispersion relation approach of Ref.~\cite{melikhov}), and the so called point form~\cite{keister,gomez}.

\item Within each of these approaches there is still a large freedom left, corresponding to the arbitrariness of the potential or of the mass operator or, more generally, to the interaction kernel. This freedom can be lifted by the choice of a general structure and shape of potentials and then by fixing the remaining parameters after comparing the model's predictions with the spectroscopical experimental data. 

\item Although we do not deny the merits of the other approaches, we would like to stress the advantages of the BT framework for the present problem. It is especially attractive in the heavy quark limit: it is covariant, it satisfies Isgur-Wise scaling and a large number of HQET sum rules which severely constrain the properties of the IW functions.~\footnote{Admittedly, this is in contrast with the finite mass situation where the good properties of the BT scheme are lost.} These sum rules do not seem to have been checked in other approaches.\par

Another, rather standard, approach used to treat  heavy-light systems is based on the Dirac equation~\cite{dirac1999,spatial}, is not relevant for the study of heavy quark currents.

\end{itemize}

Furthermore, one must stress the importance of selecting potential models (or the mass operator) by testing them over a large range of spectroscopic states, and this is what has been done in this model, covering a large range of hadronic states and with success. In that respect the Godfrey and Isgur (GI) spectroscopic hamiltonian is the most suited when working with mesons~\cite{godfrey}.

\vskip 0.2cm

The main conclusion of our calculations in the heavy quark limit of the BT approach is that the slope $\frac{d}{dw}\xi^{(1)}(1)$ is very small while the one to the second radial excitation, 
$\frac{d}{dw}\xi^{(2)}(1)$, is large. This is in contrast with the non-relativistic situation.

One should emphasize, however, that the properties of the radial excitations are usually found to be more model dependent than of the ground states. This is related to the presence of nodes 
in the wave functions. It seems then advisable to vary the wave functions within the BT framework as much as possible and check if some conclusions remain stable. 
After doing so we conclude that a general pattern is stable and that the transition to the first radial excitation remains very small for confining potentials. 

\end{enumerate}

\vskip 0.2 cm

\section{The HQET sum rule approach for the $\xi^{(n)}(w)$ Isgur-Wise function to radial excitations} \label{sectionSR}

One particularly relevant sum rule has been obtained in HQET concerning the IW functions for transitions from the ground state to radial excitations \cite{SRradialHQET}.
  
We write the sum rule as a constraint for the sum of the squared moduli of the slopes of the various radial excitations, $\frac{d} {dw } \xi^{(n)}(w)\Big|_{w=1}\equiv \frac{d} {dw } \xi^{(n)}(1)$, 
\begin{equation} \label{SR}
\sum_{n=1}^{\infty} \left|\frac{d} {dw } \xi^{(n)}(1)\right|^2 = \frac{5}{3} \sigma^2-\left(\frac{4}{3}\rho^2+\rho^4\right)\,,
\end{equation}
where $\rho^2=-\frac{d} {dw }\xi^{(0)}(1)$ is the slope of IW function, and $\sigma^2=\frac{d^2} {dw^2}\xi^{(0)}(1)$ its curvature. This sum rule gives a bound on the squared slopes of IW functions to radial excitations.
Such a sum rule is quite analogous to the Bjorken sum rule, and stands on the same rigor.  
 
We stress that it is quite useful to consider models in the infinite mass limit, independently of the precise phenomenological relevance of this limit, because important QCD-theoretical statements should be satisfied in this limit. Such a statement is Eq.~\eqref{SR}, the right hand side (r.h.s.) of which is not a priori known. It is a combination of quantities relative to the ground state, which should be first evaluated either theoretically or inferred from experiment.
In practice, one finds that there is a strong cancellation between two terms on the r.h.s. which leads to a rather strong model dependence. One can use the sum rule of Eq.~\eqref{SR} in two ways:

\begin{enumerate}

\item Deduce an approximate phenomenological bound on the radial excitations

In order to obtain a rough model-independent bound one can rely on the experimental values of $\rho^2$ and $\sigma^2$ by
 identifying
$\xi(w)$ as the ratio $G(w)/G(1)$ in $B \to D \ell \nu$. This amounts to neglecting the ${\cal O}(1/m_{c,b})$ corrections. 
Using the expansion of Ref.~\cite{caprini}, we write~\footnote{Notice that in Ref.~\cite{caprini} the expansion parameter is 
$z=(\sqrt{w+1}-\sqrt{2})/(\sqrt{w+1}+\sqrt{2})$ which we translate into expansion of $G(w)/G(1)$ in $(w-1)$.}
\begin{equation}
\frac{G(w)}{G(1)}=1-\rho_D^2 (w-1)+ \frac{1}{2} \left(\frac {67 \rho_D^2-10}{32}\right) (w-1)^2+...
\end{equation}
and therefore, 
\begin{equation}
\sigma^2=\frac {67 \rho_D^2-10} {32}\,.
\end{equation}
Numerically, the experimental fit gives~\cite{BDmunu}:
\begin{eqnarray}
\rho^2=\rho_D^2\simeq 1.19 ,\qquad \sigma^2 \simeq 2.18 \quad \Rightarrow \quad
\frac{5}{3}~ \sigma^2-\frac{4}{3}~ \rho^2-\rho^4 \simeq 0.63\,,
\end{eqnarray}
which then implies:
\begin{equation} \label{bound1}
\left| \frac{d} {dw }\xi^{(n)}(1)\right| \leq 0.8 \ .
\end{equation}
\vskip0.2cm
Notice that this is clearly below the number given in Ref.~\cite{galkin1}, and that it does not constrain any $|\frac{d} {dw }\xi^{(n)}(1)|$ to be very small. The magnitude of the bound for the slope is in fact close to the ground state slope itself, $|\frac{d} {dw }\xi^{(0)}(1)| \simeq 1$. From this alone, however, one cannot tell which are the large transitions.

\item Sum rule as a consistency condition of theoretical estimates

Using a given theoretical expression for $\xi(w)$, one can also infer a bound on the possible value of the slope for radial excitations. As given below, in the BT approach, and relying on the GI Hamiltonian as mass operator, one finds for the bound [r.h.s. of the sum rule~\eqref{SR}]:
\begin{equation}\label{bound2}
\frac{5}{3} \sigma^2-\left(\frac{4}{3}\rho^2+\rho^4\right)\simeq 0.22\,.
\end{equation}

In Ref.~\cite{SRradialBT} we demonstrated the general validity of the sum rule in the BT approach which we also check numerically below.  

The bound \eqref{bound2} is now much lower than in Eq.~(\ref{bound1}) but still much larger than the contribution we find below, arising from the first radial excitation. 
One may also wonder how one could accommodate very small values and saturate the sum. In our model, the answer is that {\em the transition to the second radial excitation is very large}, providing the main part of the sum.
\end{enumerate}
\vskip0.5cm

\section{BT results in the heavy quark limit}

We now pass onto our own calculation of IW functions in the BT scheme.

\subsection{General set up}
We consider $\xi^{(n)}(w)$, where $w=v.v'$, and $v$ ($v'$) stands for the four-velocity if the initial (final) meson state. We consider both states to be $J^P=0^-$ [$j^P=(1/2)^-$] and by denoting 
the light quark mass as $m$ and $\sqrt{p^2+m^2}$ as $p^0$, the expression reads:
\begin{eqnarray}
\xi^{(n_i,n_f)}(w)=\frac{2}{1+w} \int d^3\vec{p}~\frac {\sqrt{(p.v')(p.v)}}{p^0}~\frac {m(w+1)+p.(v+v')}{2 \sqrt{(p.v+m)(p.v'+m)}}\nonumber \\ \varphi_{n_f}(\sqrt{(p.v')^2-m^2})^* \varphi_{n_i}(\sqrt{(p.v)^2-m^2}),
\end{eqnarray}
where the radial wave functions $\varphi_{n}(p)$ are labelled by the excitation number $n$ ($n=0$ for the ground state) and they are normalized according to,~\footnote{For an $S$-wave state  we have $\int d^3 p \rightarrow 4 \pi \int p^2 dp$.}
\begin{equation}
4 \pi \int p^2 dp  \mid \varphi_{n}(p)\mid ^2=1\,.
\end{equation}

Factors of $2$ have been maintained in the expression to make manifest the normalisation to $1$ at $w=1$ for $n_i=n_f$. The first factor under the integral corresponds to the Jacobian of the Lorentz transformation of spatial wave functions, and the second factor to the Wigner rotations. We shall deal below with $n_i=0$, i.e. transitions from the ground state.

\subsection{Numerical results with the Godfrey-Isgur rest frame Hamiltonian} 
\vskip 0.5cm
Let us emphasize that while using the wave functions from the GI Hamiltonian at rest, our
treatment of the relativistic center-of-mass motion of hadrons is different from the various treatments proposed by the Toronto group (see e.g. discussion in Ref.~\cite{godfrey}). 
  
\subsubsection{Ground state results for slope and curvature and evaluation of the bound (r.h.s. of the sum rule)}

One needs the ground state IW function, in order to evaluate the bound. This means that the value of the bound depends on the model itself, 
and it will not be universal. Nevertheless, it helps verifying the consistency of the model calculation. 

As mentioned above, the r.h.s. of Eq.~\eqref{SR} is very sensitive to the values of $\rho^2$ and $\sigma^2$, which is why one needs a particularly safe and accurate calculation of 
$\rho^2$ and $\sigma^2$. Our calculations are performed by diagonalizing the Hamiltonian on a harmonic oscillator basis. We then study the dependence of the results on the dimension of the basis, from dimensions 16 to 40.
The problem is especially significant for the curvature, which involves in the integrand a higher order derivatives of the IW function.
We find a reasonable stability of the results for $\rho^2$ and $\sigma^2$ if we use formulae which minimize the order of the derivatives by deriving both the initial and the final state. 
The one for $\frac{d^2}{dw^2}\xi^{(n)}(1)$ is too long to be presented in the main text and it is given in the Appendix. The expression for $\frac{d}{dw}\xi^{(n)}(1)$ is simpler. We generalize it to different initial ($i$) and final ($f$) spatial wave functions to be applied also for transitions to radial excitations:

\begin{align}\label{eq:IWBT}
\frac{d}{dw}\xi^{(n_i,n_f)}(1)=  -4\pi\int_0^\infty dp& \biggl(\frac{1}{3} \frac{d}{dp} [p\varphi_{f}(p)]\, (p^0)^2\, \frac{d}{dp}[p\varphi_{i}(p)] \biggr. \nonumber\\
&\biggl. + \frac{1} {12}[p\varphi_{f}(p)]
\, (8-\frac{4m}{p^0+m}-\frac{m^2}{(p^0)^2})\, [p\varphi_{i}(p)] \biggr) ,
\end{align}

\noi while another expression is found below, cf. Eq.~\eqref{fullRel}.
The following values are obtained with a basis of 35 elements:
\begin{equation}\label{SRnum}
\rho^2=1.0233,\quad \sigma^2=\frac{d^2} {dw^2}\xi^{(0)}(1) =1.5810 \quad \Rightarrow\quad \frac{5}{3} \sigma^2-\left(\frac{4}{3}\rho^2+\rho^4\right)=0.2236,
\end{equation}
indeed small with respect to the phenomenological value given in Eq.~\eqref{bound1}. Further enlarging the basis leads to stable results, i.e. fully consistent with the ones quoted above. 
\vskip 0.5cm

\subsubsection{Transitions to radial excitations}

For the radial excitations, no similar accuracy is necessary. Notice also that the signs of derivatives are irrelevant since the relative phase of states is arbitrary.
\vskip 0.2cm
1) First radial excitation: With the GI mass operator, the slope of the IW function to the first radial excitation is small. 
We get
\begin{equation}\label{eq:1a}
\frac{d} {dw }\xi^{(1)}(1) \simeq -0.0088\, ,
\end{equation}
which leads to a negligible semileptonic branching ratio in the heavy quark limit (see Sec.~\ref{decay} for a discussion of the finite mass case).
\vskip 0.2cm

2) Second radial excitation: The second remarkable fact is that the slope of the IW function to the second radial excitation is large,
\begin{equation}
\frac{d} {dw }\xi^{(2)}(1)\simeq 0.4533\,.
\end{equation}
Its contribution to the semileptonic rate remains small, with respect to the ground state, due to the phase space suppression. 
It should be, however, less suppressed in the related but simpler mode, $B^0 \to D^{**} \pi$, with both $D^{**}$ and $\pi$ being electrically charged.
\vskip 0.2cm

3) Higher radial excitations $n>2$: The slopes of IW functions to higher radial excitations are again very small, even if not as small as the result given in Eq.~\eqref{eq:1a}. Notice also that they do not decrease very fast for $n>3$. We obtain: 
\begin{equation}
\frac{d} {dw }\xi^{(3)}(1)\simeq-0.071,\qquad 
\frac{d} {dw }\xi^{(4)}(1)\simeq+0.087\,.
\end{equation}

\subsubsection{The saturation of the sum rule with the GI mass operator}
\vskip 0.2cm

The above estimates of the derivatives of IW functions to radial excitations at zero recoil agree well with the expected sum of their moduli squared. In our BT approach with GI Hamiltonian the l.h.s. of Eq.~(\ref{SR}) is found to be $0.2236$, cf. Eq.~\eqref{SRnum}. Since $|\frac{d}{dw}\xi^{(2)}(1)|^2=0.2055$, the second radial excitation makes about $90\%$ of this sum and the saturation is almost complete with a few more excitations beyond the second one:
\begin{equation}
\sum_{n=3}^{n=6}\left|\frac{d} {dw }\xi^{(n)}(1)\right|^2 \simeq 0.0149\,.
\end{equation}
The whole set $n=1,..,6$ gives a sum $0.2205$, to be compared with $0.2236$. 
By further enlarging the basis of wave functions, the same sum becomes $0.2216$, thus the saturation becomes even better and one in fact may worry that 
the inclusion of further excitations might violate the bound. Indeed, for a fixed dimension of the basis of wave functions, and by increasing 
the number of terms in the sum, the sum, after decreasing with the degree of excitation, reaches a minimum, and then increases again when 
the excitation number approaches the dimension of the basis. We believe this to be an artefact due to the use of a finite dimension basis, 
as well as to the limitation of precision in the numerical calculation. 
\vskip 0.2cm

\subsection{Stability with respect to variation of the wave functions: two other examples}
\vskip 0.5cm
Based on our previous experience, the GI mass operator should be preferred to other possible choices because of its success when confronted to experimental results.  
One should, however, remain cautious as the above results involve radial excitations and, by using different sets of (reasonable) wave functions within the BT approach, check 
the robustness of the two important conclusions drawn so, namely (i) smallness of the slope of IW function to the first radial excitation, and (ii) a large size of the one to the second radial excitation. 
\vskip 0.5cm

\subsubsection{Harmonic oscillator wave functions} \label{HO}
The simplest check of the above conclusion can be made by using the basis of harmonic oscillator (HO) wave functions. For $\beta=0.5~$GeV, close to the optimal value to describe the ground state ($\beta=0.57~$GeV), we get:

- Slopes of IW functions, 
\begin{align}\label{result:GI}
&\frac{d} {dw }\xi^{(0)}(1)=-\rho^2=-1.2367\,, \nonumber \\
&\frac{d} {dw }\xi^{(1)}(1) \simeq 0.049\,, \nonumber \\
&\frac{d} {dw }\xi^{(2)}(1)\simeq0.928\,, \\
&\frac{d} {dw }\xi^{(3)}(1)\simeq-0.010\,, \nonumber \\
&\frac{d} {dw }\xi^{(4)}(1)\simeq-0.007 \,.\nonumber
\end{align}

- Curvature of the ground state,
\begin{equation}
\sigma^2=\frac{d^2} {dw^2}\xi^{(0)}(1)\simeq 2.431\,.
\end{equation}

- Saturation of the sum rule~(\ref{SR}) in the HO basis,

\begin{equation}
\mathrm{r.h.s}\quad \frac{5}{3} \sigma^2-\left(\frac{4}{3}\rho^2+\rho^4\right)\simeq 0.873\,, \qquad 
\mathrm{l.h.s.}\quad \sum_{n=1}^8 \left|\frac{d} {dw }\xi^{(n)}(1)\right|^2\simeq 0.865 ,
\end{equation}
which is quite satisfactory.
\vskip 0.5cm

Notice in particular that the dominance of the second radial excitation, already found for the GI mass operator, is even more pronounced in this case. 

\vskip 0.5cm

\subsubsection{Pseudocoulombic wave functions}
\vskip 0.2cm
The pseudocoulombic (PC) wave functions form an orthonormal set that it is particularly well suited for semirelativistic hamiltonians with square root kinetic energy, $\sqrt{k^2+m^2}$ and to relativistic equations with a linear confining potential~\cite{lucha}. A common feature they share with coulombic wave functions is an exponential falloff in $r$, but a major difference is that the coefficient in the exponential is the same for all the elements of the basis, and there is no continuous spectrum. In that latter sense it is similar in structure to the HO basis, with a general exponential factor instead of a Gaussian one. We have indeed found that the ground state wave function for the GI mass operator is exponential in $r$, which roughly corresponds to the IW function of shape $(\frac {2} {w+1})^2$~\cite{slope}. The effect of the mass is small and the dipole form is exact for $m = 0$. More precisely, one finds that numerically $\psi_0(r) \propto \exp(-0.75\,r)$, with $r$ in GeV$^{-1}$.

The PC set can be written in a simple way in the ${r}$-space :
\begin{equation}
\psi(l,n,r)=\sqrt{\frac {8 n!} {(n+l+2)!}} \, \alpha \, (2\alpha r)^l\, \exp(-\alpha r) ~L_n^{2l+2}(2\alpha r)\,,
\end{equation}
where, again, $n$ labels the radial quantum number, and $L_n^m$ stands for the Laguerre polynomial. 
As an example, for $\alpha=1~$GeV, we obtain:

- Slopes of IW functions, 

\begin{align}
&\frac{d} {dw }\xi^{(0)}(1)=-\rho^2=-0.980\,, \nonumber \\
&\frac{d} {dw }\xi^{(1)}(1) \simeq 0.062\,, \nonumber \\
&\frac{d} {dw }\xi^{(2)}(1)\simeq 0.382\,, \\
&\frac{d} {dw }\xi^{(3)}(1)\simeq -0.006\,, \nonumber \\
&\frac{d} {dw }\xi^{(4)}(1)\simeq -0.002 \,,\nonumber
\end{align}
thus very close to the results obtained by using the GI wave functions. 

- Curvature of the ground state,
\begin{equation}
\sigma^2=\frac{d^2} {dw^2}\xi^{(0)}(1)\simeq 1.452\,.
\end{equation}

- Saturation of the sum rule~(\ref{SR}) in the PC basis,
\begin{equation}
\mathrm{r.h.s}\quad \frac{5}{3} \sigma^2-\left(\frac{4}{3}\rho^2+\rho^4\right)\simeq 0.150\,, \qquad 
\mathrm{l.h.s.}\quad \sum_{n=1}^8 \left|\frac{d} {dw }\xi^{(n)}(1)\right|^2\simeq 0.150 ,
\end{equation}
which is even better than in the previous cases.
\vskip 0.5cm

\subsection{A counterexample: the set of coulombic wave functions} 
\vskip 0.5cm
If we use the coulombic wave functions and choose $\alpha=0.75~$GeV, in order to get a slope for the ground state IW function similar to the previous sets, we obtain:
\begin{align}
&\frac{d} {dw }\xi^{(0)}(1)=-\rho^2=-0.978\,, \nonumber \\
&\frac{d} {dw }\xi^{(1)}(1) \simeq 0.200\,, \\
&\frac{d} {dw }\xi^{(2)}(1)\simeq 0.018\,.  \nonumber
\end{align}
Although the results for the ground state are very similar, the pattern of slopes of IW functions to the radial excitations is completely different. 
In this case the first radial excitation actually dominates, while the second and higher ones are very small.

The natural interpretation of this qualitative difference is that the coulombic wave functions do not correspond to confining potential unlike the other choices discussed above. 

For completeness, we check the sum rule~(\ref{SR}) also in this case. We get:
\begin{equation}
\sigma^2 \simeq 1.453\, \qquad \Rightarrow\qquad \frac{5}{3} \sigma^2-\left(\frac{4}{3}\rho^2+\rho^4\right)\simeq 0.160, 
\end{equation}
and the saturation is difficult to achieve: the contribution corresponding to the first radial excitation, although now dominant, is very far from saturating the bound.\par 
In fact, the sum rule does not make sense due to the unconfined continuum spectrum. Indeed, we are studying a sum rule of the heavy quark limit of QCD at the leading order, and HQET assumes and includes the confinement mechanism.

\subsection{Conclusion on the various sets of wave functions with confined spectrum}

From the above results, one conclusion appears to be independent on the details. The three cases corresponding to a confining potential, with a discrete spectrum rising to infinity, lead to the same pattern of slopes $\frac {d} {dw} \xi^{(n)}(1)$, ($n\geq 1$), strongly dominated by the one corresponding to the second radial excitation, with all the others being very small. 
This feature deserves an explanation which we propose below, cf. Sec.~\ref{analysis}.

Let us note that this extreme dominance of the second radial excitation seems specific to the relativistic BT approach. 
It is not apparently present in the approach of Ref.~\cite{galkin1} which rather predicts a large transition to the first excitation with a similar potential. 

\subsection{Contrast with non-relativistic calculations}

Let us now stress that the observed effect is relativistic. The opposite happens in the non-relativistic (NR) calculation in which the slope of the IW function to the first radial excitation dominates the sum. To illustrate this important point we use a set of HO wave functions
where all excitations are suppressed except for the first one. The NR approximation can be obtained by taking $p\ll m$ in the BT expressions~\eqref{eq:IWBT}. We use the same parameter for numerical evaluation, $\beta=0.5$~GeV, and a large difference already appears in $\rho^2$, namely,
\begin{equation}\label{NRterm}
\rho^2_\mathrm{NR}=0.097\qquad {\rm vs.} \qquad \rho^2_\mathrm{rel}\simeq1.237,
\end{equation}
where index ``rel" indicates the full relativistic result~\eqref{result:GI}. 
Similar comparison of the NR results with those obtained in the BT scheme gives, 
\begin{align}
\frac{d} {dw }\xi^{(1)}(1)_\mathrm{NR} \simeq -0.08,\qquad &\frac{d} {dw }\xi^{(1)}(1)_\mathrm{BT} \simeq 0.05,
\nonumber\\\frac{d} {dw }\xi^{(2)}(1)_\mathrm{NR} \simeq 0,\qquad &\frac{d} {dw }\xi^{(2)}(1)_\mathrm{BT} \simeq 0.93.
\end{align}
In other words the relativistic effects are huge, which is expected since one is in a very relativistic regime for the light quark, namely $\beta/m \gtrsim 1$. Indeed, with  $\varphi \propto \exp(-\beta^2 r^2/2)$, $\beta=0.57~$GeV and $m=0.22~$GeV, one obtains $\beta/m \in (2,  3)$.

Relativistic effects are not always so large for all physical quantities. The above example is specific to the slopes. Moreover, it must be recalled that to perform a fair comparison between non-relativistic and relativistic approaches, it would be convenient to rescale the parameters of our NR version so as to describe the same spectrum. For instance, this procedure would induce a change in the mass of the quark, which is usually found around $0.33~$GeV in NR case, instead of $0.22~$GeV for the GI mass operator.
The inverse radius $\beta$ would have to be changed from $\beta=0.5~$GeV to a smaller value, depending on the potential. In fact, one can get an idea of the final result for the $\rho^2_\mathrm{NR}$ from the results tabulated in Ref.~\cite{slope}. The so-called ISGW model \cite{ISGW}, using a Schr\"oedinger wave equation and $m \simeq 0.3$~GeV, leads to a reasonable spectrum but it yields $\rho^2 \simeq 0.33$ 
if calculated in the NR case~\cite{slope}. That value is much larger value than the one in Eq.~(\ref{NRterm}) but still much smaller than the required $\rho^2 \simeq 1$. Only the relativistic corrections of the BT treatment of the center of motion can fill the gap. In that way, the ISGW spectroscopic model results in $\rho^2 \simeq 1.28$~\cite{slope}.

In any case, one thing remains clear: whatever the parameters of the NR model with HO wave functions, 
the slope at origin for the transition to the first radial excitation is the only one which does not vanish, in complete contrast with the relativistic case. 
This can also be seen from the expression for the derivative of the IW function in the NR case, $\frac{d}{dw}\xi^{(n_i,n_f)}(1)=- (1/3) \left<f|r^2|i \right>$. 
Since a HO radial excitation wave function ($n>0$) is a polynomial times the ground state one, with degree $2n$, it is clear that the state $r^2|i\rangle$ is a linear superposition of the ground state and the first radial excitation, and it is orthogonal to the states with $n>1$. 

Therefore, there is something qualitatively new in the relativistic calculation, i.e.  a dominant higher excitation $n=2$. We now try to interpret this result.

\subsection{Relation between the relativistic corrections for center-of-mass motion and the pattern of transitions to radial excitations} 
\label{analysis}

One can find a simple relation between this remarkable pattern of transitions to radial excitations and the structure of relativistic corrections 
for center-of-mass motion of the hadrons in the BT approach.

Let us recall that the relativistic effect in our calculation is compound of several ones. In the case of the slope $\rho^2$ they arise from~\cite{slope,duality}:

\begin{enumerate}
\item {\em Relativistic effects in the mass operator.} The relativistic GI mass operator, describing the states at rest, with a relativistic kinetic energy and other relativistic effects affecting the potential are contained in the internal wave functions and in the hadron masses which are used in a concrete calculation.

\item {\em The effects coming from the boost of the states from their rest frame.} These are the specific contributions from the BT approach. They are displayed explicitly in the expression for $\rho^2$ in terms of the internal wave functions. They include both the purely spatial effects (as would be present for scalar quarks) and the quark spin effects (e.g. Wigner rotations).  

\end{enumerate}
We decompose the effects of center-of-mass motion into contributions having increasing power of the internal velocity $(v/c)^n$, where $v/c$ represents, say, $\left< k\right> /m$, with $k$ being a generic internal momentum.
We further re-express these effects in a form in which the derivatives with respect to internal momentum always affect only the initial ground state wave function $\varphi_i$. In other words we write the slope as a sum of matrix elements over the internal wave functions $\left<f\mid {\cal O}\mid i\right>$, ${\cal O}$ being a local operator in internal momentum space. This is different from the former choice which was conceived for numerical efficiency. We are now motivated by the need of a transparent interpretation of the various contributions.

Starting from the expression for $\rho^2$ given in Ref.~\cite{duality}, we have:
\begin{equation} \label{standard}
\frac {d}{dw}\xi^{(n)}(1)=-\left<n \left| \left(\frac {k^0 z +z k^0} {2}\right)^2 \right| 0\right> -\frac{1}{4} \delta_{0,n} -\frac{1}{6} \left<n \left| \left(\frac {k^2} {k^0+m}\right)^2 \right| 0\right>,
\end{equation}
where the average is over {\em internal} wave functions and $z$ is an arbitrary direction. The matrix element exhibits rotational symmetry, and $k^0=\sqrt{k^2+m^2}$.
The last two terms come from the spin of the active heavy quark and from the spin of the light spectator quark (Wigner rotation).~\footnote{We consider the vector current to define $\xi$, the matrix element of $j^0$ with non-relativistic normalization, and a factor for the active quark $\bar{u}_{{s'_1}} \gamma_{0} u_{s_1}$ with non-relativistic normalization of spinors $u^{\dagger} u=1$. The final $\xi$ is of course independent of the choice of the current.} The first term,
\begin{equation}
-\left<n\left| \left(\frac {k^0 z +z k^0} {2}\right)^2 \right| 0\right>\,,
\end{equation}
is then the purely spatial effect, which would be present already for scalar quarks. It comes from the Lorentz transformation of the spatial internal wave function.
It can be reexpressed in terms of derivatives with respect to the momentum. Using rotational symmetry and denoting $|\vec{k}|=k$, one can write:
\begin{align}\label{spatial}
-\left< n\left|  \left(\frac{k^0 z +z k^0} {2}\right)^2 \right| 0\right> 
= &\frac{1}{3}\ m^2 \int d\vec{k}~\varphi_f(\vec{k})^\ast \left(\frac {2}{k} \frac{d} {dk}+\frac{d^2} {dk^2}\right)  \varphi_i(\vec{k})
\nonumber\\
&+ \frac{1}{3} \int d\vec{k}~k^2~\varphi_f(\vec{k})^* \left(\frac {4}{k} \frac{d} {dk}+\frac{d^2} {dk^2}\right)  \varphi_i(\vec{k})\nonumber\\
&+ \frac{1}{2} \int d\vec{k}~\varphi_f(\vec{k})^* \varphi_i(\vec{k})-\frac{1}{12} \int d\vec{k}~\frac{k^2}{k_0^2}\varphi_f(\vec{k})^* \varphi_i(\vec{k})\,.
\end{align}
Let us count the powers in $v/c$ in various terms of \eqref{standard}, and first in \eqref{spatial}. We are not performing an expansion, we are just estimating the order of various terms whose sum represents the exact result.

The first line in \eqref{spatial} is the lowest order, i.e. the non-relativistic expression, of order $(v/c)^{-2}$. The quantity under the integral is obviously $\mathcal{O}(k^{-2})$:
\begin{equation}
\frac{1}{3}\ m^2 \int d\vec{k}~\varphi_f(\vec{k})^\ast \left(\frac {2}{k} \frac{d} {dk} +\frac{d^2} {dk^2}\right)  \varphi_i(\vec{k})\,.
\end{equation}

We now consider relativistic effects, which are of higher order in $v/c$: of the order $(v/c)^{0}$,
\begin{eqnarray}
\frac{1}{3} \int d\vec{k}~k^2~\varphi_f(\vec{k})^* \left(\frac {4}{k} \frac{d} {dk}+\frac{d^2} {dk^2}\right)  \varphi_i(\vec{k})+
\frac{1}{2} \int d\vec{k}~\varphi_f(\vec{k})^* \varphi_i(\vec{k})\,,
\end{eqnarray}
and of the order $(v/c)^{2}$,
\begin{equation}
-\frac{1}{12} \int d\vec{k}~\frac{k^2}{k_0^2}\varphi_f(\vec{k})^* \varphi_i(\vec{k})\,.
\end{equation}
These correspond to the effect of the Lorentz transformation on the spatial wave functions $\varphi_{i,f}(\vec{k})$.

For spin-$1/2$ quarks, the result given in Eq.~(\ref{standard}) includes two additional terms coming from 
(i) the free heavy quark current, and (ii) the light quark Wigner rotations (the last two terms). One then ends up with:
\begin{align} \label{fullRel}
\frac{d}{dw}\xi^{(n_i,n_f)}(1)=&\frac{1}{3} m^2 \int d\vec{k}~\varphi_f(\vec{k})^* \left(\frac {2}{k} \frac{d} {dk}+\frac{d^2} {dk^2}\right)  \varphi_i(\vec{k})\nonumber \\
&+\frac{1}{3} \int d\vec{k}~k^2~\varphi_f(\vec{k})^* \left(\frac {4}{k} \frac{d} {dk}+\frac{d^2} {dk^2}\right)  \varphi_i(\vec{k})
+\frac{1}{4} \int d\vec{k}~\varphi_f(\vec{k})^* \varphi_i(\vec{k})\nonumber \\
&-\frac{1}{12} \int d\vec{k}~\frac{k^2}{k_0^2}\varphi_f(\vec{k})^* \varphi_i(\vec{k})-\frac {1}{6}\int d\vec{k}~\frac{k^2}{(m_q+k_0)^2}\varphi_f(\vec{k})^* \varphi_i(\vec{k})\,.
\end{align}
The coefficient of the term in $\int d\vec{k}~\varphi_f(\vec{k})^* \varphi_i(\vec{k})$~has changed from $1/2$ to $1/4$ by including the contribution -$1/4$ of the heavy quark coming from the current matrix element $\bar{u}_{s_1'} \gamma_{\mu} u_{s_1}$, and the Wigner rotation effect is the last term in Eq.~(\ref{fullRel}). 

The three lines of Eq.~\eqref{fullRel} represent now respectively the NR contribution $\mathcal{O}[(v/c)^{-2}]$ (I), the $\mathcal{O}[(v/c)^{0}]$ terms (II), and a $\mathcal{O}[(v/c)^{2}]$ 
relativistic correction (III). To discuss the magnitude of the three contributions, one may for instance use a set of HO wave functions. It appears then that I is of $\mathcal{O}(m^2 R^2)$; II is most remarkably  independent of $m$ and of $R^2$. In other words, II is scale independent, it does not depend on the size of the bound state. III is $\mathcal{O}[1/(m^2 R^2)]$. In a non-relativistic system $m^2 R^2 \gg 1$ implying a hierarchy I $\gg$ II $\gg$ III. With the realistic parameters, $m^2 R^2 < 1$, one finds quite a different pattern with the following striking results for a large range of $m^2 R^2$, $m=0.22$,  $1 <R^2<10$, a range which includes the realistic values:

-- the total result for $\frac{d}{dw}\xi^{(n)}(1)$ is close to $1$ for $n=0,2$, and small for $n=1$ or $n>2$; 

-- II, when different from zero, is by far the dominant term (i.e. for $n=0,2$), the NR contribution I being reasonably small;

-- for the elastic transition, the value of the term II is fixed to $-1$, i.e. it is independent of 
$R^2$ or $\beta^2=1/R^2$. For the transition to the first radial excitation, instead, it is always $0$, as well as for the transitions to $n>2$; finally, for $n=2$, it is always $\simeq 0.913$;

-- on the whole, the spin effects are rather small, and the total result is close to what it would be for scalar quarks, and its main contribution is the term coming from the Lorentz transformation 
of the spatial part of the wave function, term II.

Qualitatively, these conclusions obtained with the set of HO wave functions extend to the Godfrey-Isgur system of wave functions.

Let us understand a little more the result for the scale independent term II.
Leaving aside $\frac{1}{4} \int d\vec{k}~\varphi_f(\vec{k})^* \varphi_i(\vec{k})$ which gives no transition, one sees that for a HO ground state for initial state, 
\begin{eqnarray}
\frac{1}{3} ~k^2~\left(\frac {4}{k} \frac{d} {dk}+\frac{d^2} {dk^2}\right)  \varphi_i(\vec{k}),
\end{eqnarray}
is an exact linear superposition of the $n=0$ and $n=2$ HO states with algebraic coefficients:
\begin{eqnarray}
\frac{1}{3} ~k^2~\left(\frac {4}{k} \frac{d} {dk}+\frac{d^2} {dk^2}\right)  \varphi_0(\vec{k})=
\sqrt{\frac{5}{6}}~\varphi_2(\vec{k}) - \frac{5}{4}~\varphi_0(\vec{k}).
\end{eqnarray}
The contribution of II to the slope of the ground state $\frac{d}{dw}\xi^{(0)}(1)$ is:
\begin{eqnarray}
-\frac{1}{3} \times \frac{15}{4}+\frac{1}{4}=-1\,,
\end{eqnarray}
as anticipated above, while its contribution to $\frac{d}{dw}\xi^{(2)}(1)$ is 
\begin{eqnarray}
\sqrt{\frac{5}{6}}\simeq 0.913\,,
\end{eqnarray}
and it does not contribute to other radial excitations. These numbers are already not far from the total results given in subsection~\ref{HO}, which include spin. This is also 
the general pattern that we find with GI or pseudocoulombic wave functions.

The general conclusion is then that near $w=1$ the term II, which corresponds to the Lorentz transformation of the spatial wave function from the rest frame, contributes almost entirely to the ground state and to the second radial excitation, with large and comparable values of the slope $\simeq 1$, while the slope of the transition to any other radial excitation, and consequently the transition itself, is very small.
This feature of the Bakamjian-Thomas model in the heavy quark limit is not found apparently in other works. However, in principle, it is  {\it a consequence of a very general phenomenon which is the Lorentz transformation of the spatial wave function}.

\section{How to check the BT prediction on the pattern of radial excitations?} \label{decay} 

\subsection{Comparisons at $m_Q \to \infty$ with lattice calculations} 

As we have stressed, we have only definite covariant predictions for the $m_Q\to\infty$ limit of the current matrix elements in the BT approach, and this is why we stuck to this limit.~\footnote{By ``current matrix elements" we always mean that the current operator is simply the sum of standard currents over all the quarks, in the spirit of the old additivity or ``impulse" approximations. Otherwise, one could always maintain covariance by the introduction of two-body current operators. We thank B.~Keister for stressing that point.} Our $m_Q\to \infty$ discussion is valuable in itself because: 1) it can be confronted with other statements made in this limit, as we have done above for the sum rules and other models; and 2) in principle, it could be also checked by lattice QCD calculations made directly in the $m_Q\to\infty$ limit, as it has been done for the transitions mediated by a light quark current~\cite{blossierDistr,LQdistributions} and previously for heavy current transitions to $L=1$ excitations~\cite{HQdistributions}, or one could compare to calculations performed at very large quark masses.
 
	The case of heavy currents poses specific problems in the exact infinite heavy quark mass limit since a finite $w \neq 1$ corresponds to infinite momentum transfer. On the lattice, one can then reach only $w=1$, where the IW function for the inelastic transition vanishes. A similar problem was encountered in transitions to $L=1$. One way to get around that difficulty is to use a derivative current operator that does not vanish at $w=1$, factorizing a trivial factor $\propto (w-1)$ (see reference in subsection \ref{methods}). This derivative expression is obtained due to the identities between the coefficients of the subleading corrections in $1/m_{b,c}$ of current type (which can indeed be expressed as covariant derivatives) and the leading IW function.

A similar solution could perhaps be devised for the transitions to the radially excited states, by a manipulation of the identities found in our discussion \cite{nous5}. For instance, one gets
\begin{eqnarray}
L_5^{(b)(n)}(1)=-L_5^{(c)(n)}(1)=\Delta E^{(n)}~\frac{d}{dw}\xi^{(n)}(1)\,,
\end{eqnarray}
where $L_5^{(b,c)(n)}(w)$ are subleading form factors, for $1/m_{b,c}$ corrections that generalize the corresponding elastic quantities defined in Ref.~\cite{FN}. Equivalently, one could extract the relevant coefficients with an infinitely heavy $b$ quark, and by varying the $c$ quark mass.
 
One could also calculate directly the current matrix element for finite momenta at finite but very large masses (using NRQCD), with the current matrix element approximating the IW function. As we said, the attempt with this last method reported in Ref.~\cite{hein} should be considered as preliminary since the extraction of excited states can be much improved nowadays. 

\subsection{Discussion of phenomenological tests} 

Of course, it would be desirable to test the above striking predictions in phenomenological $b\to c$ processes but several problems seem to render this goal difficult. 
Such tests would in principle require full finite mass calculations in the quark model. As we already emphasized it in the introduction, we do not trust the predictions of the BT approach at finite mass, specifically for inelastic transitions, while in the elastic case they seem satisfactory. We have explained the difficulty in detail for the case of transitions between the $L=0$ ground state and the $L=1$ orbitally excited states in Ref.~\cite{dong}: certain identities at order $1/m_Q$ are not satisfied in the quark model approach. In another recent paper~\cite{nous5}, we presented similar (general) identities for the transitions to radial excitations, and found a similar conclusion: one identity at order $1/m_Q$ cannot be satisfied in the inelastic case, while it is satisfied in the elastic case of BT with finite quark masses. 

Indeed, in the transitions to $L=1$, we have found for the non-leptonic decays that the finite mass calculation gives a much too large rate for the measured decay to $0^+$ $D$-meson, in contradiction with experiment~\cite{dong}, and this leaves us in theoretical incertitude.

A recipe has been successful in the case of the transitions between $L=0$ and $L=1$, namely to make predictions by combining the heavy quark mass limit of the amplitudes with a realistic phase space. 
Indeed, in this way, in the non-leptonic decays, applying factorization, one gets reasonable agreement for the ``elastic" decay $B \to D^{(*)}\pi$ and for the now well measured decays $B \to D(2^{+},0^{+})\pi$~\cite{noteRoudeau}. 
One is then encouraged to try and apply this second approach to the $B \to D^{(n)(*)}$ transitions $(n=1, 2)$.
In the $B \to D^{(1)(*)}$ case, the $m_Q\to \infty$ result is so small that any $1/m_Q$ correction will destroy the initial result.  On the other hand, this way of estimating the physical processes could be more trustable for $B \to D^{(2)(*)}$ because the amplitude is large. 

All in all, the best tests should be through lattice QCD near the $m_Q\to \infty$ limit.

\section{Conclusion} 

One must stress first the remarkable feature of the pattern of transitions to the radial excitations in the BT approach at $m_Q \to \infty$: these transitions are very strongly dominated by the transition to the second excitation, while $n=1$ and $n>2$ excitations have a very small transition matrix element. Our approach has had some notable successes for current matrix elements, but it is important that its conclusions are checked in some way. There seems to be no really easy way for such a test. All in all, the best one could do would be a lattice QCD calculation in the limit $m_Q\to \infty$, using suitable operators to allow the calculation at $w=1$, or by using the finite but large masses.
 
 \vspace*{2.7cm}
 
\section*{Acknowledgements} This is the last paper in which our late colleague Jean-Claude Raynal has participated till his sudden death. After so many years of very fruitful collaboration, we would like to say that we will not forget him. We thank Beno\^it Blossier for very useful discussions.

\newpage

\section*{Appendix: Expressions for the numerical calculation of the slope and the curvature}

To check the sum rules, one requires high accuracy in the calculation of slopes and  curvatures. We have found that the following expressions are well adapted for these numerical calculations:
\begin{align}
\frac{d}{dw}\xi^{(n_i,n_f)}(1) \equiv -\rho^2=& -4\pi \int_0^\infty dp\, \bigg\{\frac{1}{3} \frac{d}{dp} [p\varphi_{f}(p)] (p^0)^2 \frac{d}{dp}[p\varphi_{i}(p)]\nonumber\\
&+ \frac{1} {12} [p\varphi_{f}(p)]~
\left(8-\frac{4m}{p^0+m}-\frac{m^2}{(p^0)^2}\right)~[p\varphi_{i}(p)]\bigg\}\,,
\end{align}

\begin{align}
\frac{d^2}{dw^2}\xi^{(n_i,n_f)}(1)  \equiv \sigma^2=& 4\pi\int_0^\infty dp \bigg\{\frac{1}{15} \frac{d^2}{dp^2} [p\varphi_{f}(p)](p^0)^4 \frac{d^2}{dp^2}[p\varphi_{i}(p)] \nonumber \\
&+\frac{1} {30}\frac{d}{dp}[p\varphi_{f}(p)] (p^0)^2 \left(12-\frac{4m}{p^0+m}-\frac{m^2}{(p^0)^2}\right)\frac{d}{dp} [p\varphi_{i}(p)]\nonumber \\   
&+ \frac{1}{240} [p\varphi_{f}(p)] \frac{1}{(p^0)^4(p^0+m)^2}\bigg(170~m^6+593~m^4 p^2+600~m^2 p^4+192~p^6 \nonumber \\
&+ 2m(101~m^4+244~m^2 p^2+128~p^4)~p^0\bigg)~[p\varphi_{i}(p)]\bigg\}\,.
\end{align}

\end{document}